# On the radiative and thermodynamic properties of the extragalactic far infrared background radiation using *COBE* FIRAS instrument data

**Anatoliy I Fisenko, Vladimir Lemberg**

*ONCFEC Inc., 250 Lake Street, Suite 909, St. Catharines, Ontario L2R 5Z4, Canada*

*E-mail: afisenko@oncfec.com*

**Abstract** Using the explicit form of the function to describe the average spectrum of the extragalactic far infrared background (FIRB) radiation measured by the *COBE* FIRAS instrument in the 0.15 – 2.4 THz frequency interval, the radiative and thermodynamic properties, such as the total emissivity, total radiation power per unit area, total energy density, number density of photons, Helmholtz free energy density, entropy density, heat capacity at constant volume, pressure, enthalpy density, and internal energy density are calculated. The calculated value of the total intensity received in the 0.15 – 2.4 THz frequency interval is 13.6 nW $m^{-2}$ $sr^{-1}$, and comprises about 19.4% of the total intensity expected from the energy released by stellar nucleosynthesis over cosmic history. The radiative and thermodynamic functions of the extragalactic far infrared background (FIRB) radiation are calculated at redshift $z$ = 1.5.

Percent

## 1 Introduction

The extragalactic infrared background radiation (EIB) spectrum represents the redshifted energy released from the first stellar objects, protogalaxies, and galaxies throughout cosmic history. This background radiation was detected by analyzing the *COBE* FIRAS and DIRBE data (Puget et al. 1996; Fixsen et al. 1998; Hauser et al. 1998). As noted by Dole et al. (2006), this detected background represents about half of the extragalactic background light (EBL), because it covers wavelength range from 1 μm to 1000 μm, i.e. the infrared region. The far infrared wavelength interval from 125 μm to 2000 μm was used in the *COBE* FIRAS experiments and their data contain foreground emission from the interplanetary, Galactic interstellar dust and extragalactic background emission (Fixsen et al. 1998). To separate the different components of the emission was used three different methods. As a result, the analytical expression for average spectrum in the $v$ = 125 – 2000 μm wavelength interval at temperature $T = 18.5 \pm 1.2\,\mathrm{K}$ has been proposed to describe the extragalactic Far InfraRed Background (FIRB) spectrum (Fixsen et al. 1998). It is essential to note that this spectrum differs from the monopole spectrum (Fixsen et al. 1994) by the spectral emissivity coefficient.

The extragalactic far infrared background spectrum was detected in the finite wavenumber range from 125 μm to 2000 μm. Thus, for the construction of thermodynamics of the extragalactic far infrared background (FIRB) radiation, as well as the radiative properties, we have to use this specific range of wavenumbers.

In (Fisenko & Lemberg 2014), the exact expressions for the temperature dependences of the thermodynamic and radiative properties of the Cosmic Microwave Background radiation for the monopole and dipole spectra were obtained. However, the extragalactic spectrum differs from the blackbody spectrum because of the spectral emissivity factor $\varepsilon(\tilde{v}, T)$ and, thus, difficult to construct the exact expressions. In this case, we will use numerical methods for calculating the thermodynamic and radiative properties of the FIRB radiation.

In previous studies (Fisenko & Ivashov 2009; Fisenko & Lemberg 2012; Fisenko & Lemberg 2013), the thermodynamics of the thermal radiation of real bodies, such as molybdenium, luminous flames, *stoichiometric* carbides of hafnium, titanium and zirconium, and $ZrB_2$-SiC-based ultra-high temperature ceramics in the finite range of frequencies at high temperatures were constructed by using numerical methods. The calculated values of the radiative and thermodynamic functions were in good agreement with experimental data.

In the present work, the values of the radiative and thermodynamic functions of the extragalactic far infrared background radiation, such as the total radiation power per unit area, total energy density, number density of photons, Helmholtz free energy density, entropy density, heat capacity at constant volume, pressure, enthalpy density, and internal energy density, received in the $(0.15\text{-}2.4)\,\mathrm{THz}$ frequency interval at the mean temperature $T$ = 18.5 K are calculated. Using formulas $\nu = \nu_0(1+z)$ and $T = T_0(1+z)$, the radiative and thermodynamic functions were calculated at redshift $z$ = 1.5. In the appendix, the values of the radiative and thermodynamic functions of the Cosmic Microwave Background radiation for the monopole and dipole spectra at redshift $z$ = 1089 are obtained.

## 2 Extragalactic Spectrum

According to (Fixsen et al. 1998), the average spectrum of the extragalactic Far InfraRed Background (FIRB) radiation in the $5\,\mathrm{cm}^{-1} \le \tilde{\nu} \le 80\,\mathrm{cm}^{-1}$ range of wavenumbers is presented as:

$$I_{\tilde{\nu}}(T) = \varepsilon(\tilde{\nu}, T) B_{\tilde{\nu}}(T)\,, \tag{1}$$

where

$$\varepsilon(\tilde{\nu}, T) = \varepsilon_0 \left( \frac{\tilde{\nu}}{\tilde{\nu}_0} \right)^{0.64 \pm 0.12} \tag{2}$$

is the spectral emissivity. Here $\varepsilon_0 = (1.3 \pm 0.4) \times 10^{-5}$, $\tilde{\nu}_0 = 100\ \mathrm{cm}^{-1}$ and $T = 18.5 \pm 1.2\,\mathrm{K}$. $B_{\tilde{\nu}}(T)$ is the Planck function in the wavenumber domain.

For the construction of the thermodynamics of FIRB radiation, using the *COBE* FIRAS instrument data, hereinafter, it is convenient to present the Planck function $B_{\tilde{\nu}}(T)$ in the frequency domain. Using the relationship

$$B_{\tilde{\nu}}(T)\mathrm{d}\tilde{\nu} = B_{\nu(\tilde{\nu})}(T)\mathrm{d}\nu\,, \tag{3}$$

where $\tilde{\nu}$ stands for wavenumber, that can be related to the frequency $\nu$ via the transformation $\nu(\tilde{\nu})$, yields then

$$B_\nu(T) = B_{\tilde{\nu}}(T)\frac{d\tilde{\nu}}{d\nu} = \frac{2h}{c^2}\frac{\nu^3}{e^{\frac{h\nu}{k_\mathrm{B}T}} - 1} \quad . \tag{4}$$

Eq. (4) is described the Planckian radiation in the frequency domain.

Using the relationship $I_\nu(T) = \frac{4\pi}{c}B_\nu(T)$, for the spectral energy density $I_\nu(T)$, we obtain (Landau & Lifshitz 1980)

$$I_\nu(T) = \frac{8\pi h}{c^3}\frac{\nu^3}{e^{\frac{h\nu}{k_\mathrm{B}T}} - 1} \quad . \tag{5}$$

According to Eq. 5, the Eq. 1 in the frequency domain $\nu$ has the structure:

$$I'_\nu(T) = \varepsilon(\nu, T) I_\nu(T) \tag{6}$$

with

$$\varepsilon(\nu, T) = \varepsilon_0 \left(\frac{\nu}{\nu_0}\right)^{0.64 \pm 0.12} , \tag{7}$$

where $\varepsilon_0 = (1.3 \pm 0.4) \times 10^{-5}$, $\nu_0 = 3\,\mathrm{THz}$ .

Then, in accordance with Eq. 6, the total energy density of the extragalactic FIRB radiation in the frequency domain has the form

$$I_0(\nu_1, \nu_2, T) = \int_{\nu_1}^{\nu_2} I'_\nu(T)\mathrm{d}\nu = \int_{\nu_1}^{\nu_2} \varepsilon(\nu, T) I_\nu(T)\mathrm{d}\nu . \tag{8}$$

The Stefan-Boltzmann law or the total radiation power per unit area received in the finite frequency range is defined as

$$I'_0(\nu_1, \nu_2, T) = \frac{4\pi}{c} I_0(\nu_1, \nu_2, T) . \tag{9}$$

The total emissivity in the finite range of frequencies $\nu_1 \le \nu \le \nu_2$ is determined as

$$\varepsilon(T) = \frac{I_0(\nu_1, \nu_2, T)}{\int_{\nu_1}^{\nu_2} I_\nu(T)\mathrm{d}\nu} \tag{10}$$

The integrals are evaluated over the spectral band $\nu_1$ and $\nu_2$ .

In the equation 10 the total energy density of the blackbody radiation according to (Fisenko & Lemberg 2014) has the following structure:

$$\int_{\nu_1}^{\nu_2} I_\nu(T)\mathrm{d}\nu = \frac{48\pi(k_\mathrm{B}T)^4}{c^3h^3}[P_3(x_1) - P_3(x_2)], \tag{11}$$

where $x = \frac{h\nu}{k_\mathrm{B}T}$. $P_3(x)$ is defined as

$$P_3(x) = \sum_{s=0}^{3} \frac{(x)^s}{s!}\mathrm{Li}_{4-s}(e^{-x}). \tag{12}$$

Here

$$\mathrm{Li}_{4-s}(e^{-x}) = \sum_{k=1}^{\infty} \frac{e^{-kx}}{k^{4-s}} \tag{13}$$

is the polylogarithm function (Abramowitz & Stegun 1972).

Then, Eq. 10 has the form

$$\varepsilon(T) = \frac{c^3h^3}{48\pi(k_\mathrm{B}T)^4} \frac{I_0(\nu_1,\nu_2,T)}{[P_3(x_1) - P_3(x_2)]}. \tag{14}$$

Knowledge of the total energy density allows us to construct the thermodynamics of FIRB radiation as follows:

1) Helmholtz free energy density $f = \frac{F}{V}$:

$$f = -\frac{1}{3} I_0(\nu_1,\nu_2,T) \tag{15}$$

2) Entropy density $s = \frac{S}{V}$:

$$s = \frac{1}{3}\frac{\partial I_0(\nu_1,\nu_2,T)}{\partial T} \tag{16}$$

3) Heat capacity of at constant volume per unit volume $c_V = \frac{C_V}{V}$:

$$c_V = \frac{T}{3}\left(\frac{\partial^2 I_0(\nu_1,\nu_2,T)}{\partial^2(T)}\right)_V \tag{17}$$

4) Pressure of photons $P$:

$$P = \frac{1}{3} I_0(v_1, v_2, T) \tag{18}$$

5) Enthalpy density $h = \frac{H}{V}$:

$$h = \frac{T}{3} \frac{\partial I_0(v_1, v_2, T)}{\partial T} \tag{19}$$

6) Internal energy density $u = \frac{U}{V}$:

$$u = -\frac{1}{3}\left( I_0(v_1, v_2, T) - T \frac{\partial I_0(v_1, v_2, T)}{\partial T} \right) \tag{20}$$

7) The number density of photons $n = \frac{N}{V}$:

$$n = \frac{8\pi h}{c^3} \int_{v_1}^{v_2} \frac{\varepsilon(v, T)\, v^2}{e^x - 1} dv \tag{21}$$

To calculate the values of the total energy density Eq. 8 and the number density of photons Eq. 21 Simpson's rule was applied (Abramowitz & Stegun 1972). Knowledge of the total energy density allows us to calculate the total emissivity, Helmholtz free energy density and pressure in the finite range of frequencies from $0.15\,\mathrm{THz}$ to $2.4\,\mathrm{THz}$ at the mean temperature $T = 18.5\,\mathrm{K}$. The results are presented in Table 1.

The values of the entropy density, heat capacity at constant volume, enthalpy density and internal energy density were calculated using the method of numerical differentiation (Abramowitz & Stegun 1972). The calculated values of these thermodynamic functions are presented in Table 1.

Let us calculate the total intensity $\tilde{I}(v_1, v_2, T)$ received in the finite frequency range $(0.15\text{-}2.4)\,\mathrm{THz}$. According to (Fixsen et al. 1998), a value of $\tilde{I}(v_1, v_2, T)$ is 14 nW m$^{-2}$sr$^{-1}$. Using the relationship between the total intensity $\tilde{I}(v_1, v_2, T)$ and the total energy density $I_0(v_1, v_2, T)$

$$\tilde{I}(v_1, v_2, T) = \frac{c}{4\pi} I_0(v_1, v_2, T), \tag{22}$$

in accordance with Table 1, we obtain $\tilde{I}(v_1, v_2, T) = 13.6\,\mathrm{nW\,m^{-2}sr^{-1}}$. This value is about 19.4% of the total expected flux (70 nW m$^{-2}$sr$^{-1}$) associated with the production of metals throughout the history on the universe (Dwek et al. 1998). As seen, the calculated value is in good agreement with the experimental result obtained in (Fixsen et al. 1998).

Here it is important to note that the calculated values of the thermodynamic and radiative functions represent the state of the universe at the present time, which corresponds to the redshift z = 0.

In (Pe, Fall & Hauser 1999), it was noted that 85% of the integrated galaxy light arises from galaxies at z < 1.5. This value corresponds to the peak in the history of star-formation and rapid growth of the stellar mass in galaxies. Therefore, we calculate the radiative and thermodynamic properties of the extragalactic FIRB radiation at redshift $z = 1.5$.

It is well known that in an expanding universe the temperature $T$ and the frequency $v$ depends on the redshift z (Sunyaev & Zel'dovich 1980), in accordance with the formulas $v = v_0(1+z)$ and $T = T_0(1+z)$. In this case, for numerical calculation of the radiative and thermodynamic functions of the extragalactic FIRB radiation for the average spectrum, we should the mean temperature $T$ = 18.5 K replace by $T$ = 46.25 K and a finite frequency interval $0.15\,\text{THz} - 2.4\,\text{THz}$ should be replaced by 2.625 THz – 42 THz. In Table 2, the calculated values of the radiative and thermodynamic functions of the extragalactic far infrared background radiation at redshift z = 1.5 are presented.

## 3 Conclusions

In this paper, the values of the radiative and thermodynamic functions of extragalactic far infrared background radiation, such as the total radiation power per unit area, total energy density, number density of photons, Helmholtz free energy density, entropy density, heat capacity at constant volume, pressure, enthalpy density, and internal energy density in the finite range of frequencies $0.15\,\text{THz} \le v \le 2.4\,\text{THz}$ at the temperature $T$ = 18.5 K are calculated. The calculated value of the total intensity $\tilde{I}(T) = 13.6\,\text{nW m}^{-2}\,\text{sr}^{-1}$ received in the finite range of frequencies from $0.15\,\text{THz}$ to $2.4\,\text{THz}$ is in good agreement with the experimental data (Fixsen et al. 1998). This value is about 19.4% of the total intensity (70 nW $\text{m}^{-2}\text{sr}^{-1}$) expected from the energy released by stellar nucleosynthesis over cosmic history (Dwek et al. 1998).

The radiative and thermodynamic functions of the extragalactic far infrared background radiation at redshift z = 1.5 are calculated.

In conclusion we note that, as in the case of the monopole and dipole spectra (Fisenko & Lemberg) is desirably to obtain exact expressions for the temperature dependences of the radiative and thermodynamic functions of the extragalactic far infrared background radiation. This topic will be point of discussion in subsequent publication.

## Appendix

In the previous paper (Fisenko & Lemberg 2014), the radiative and thermodynamic state functions of the Cosmic Microwave Background radiation for the monopole and dipole spectra were calculated at redshift $z$ =1089. There was used the finite range of frequencies from $v_1 = 0{,}0499$ PHz to $v_2 = 0{,}499$ PHz . However, in an expanding universe the frequency $v$ depends on the redshift z (Sunyaev & Zel'dovich, 1980), in accordance with the formula $v = v_0(1+z)$. Taking into account this fact, we recalculate the radiative and thermodynamic functions using the following 65.4 – 654 THz frequency range. In Table 3, the radiative and thermodynamic functions at redshift $z$ = 1089 are presented.

| Quantity | Extragalactic Average Spectrum $\nu_1 \le \nu \le \nu_2$ |
|---|---|
| $I_0(\nu_1, \nu_2, T)$ $[\mathrm{J\ m^{-3}}]$ | $5.6839\times10^{-16}$ |
| $I_0'(\nu_1, \nu_2, T)$ $[\mathrm{W\ m^{-2}}]$ | $4.2601\times10^{-8}$ |
| $\varepsilon(T)$ | $7.3308\times10^{-6}$ |
| $f\ [\mathrm{J\ m^{-3}}]$ | $-1.8946\times10^{-16}$ |
| $u\ [\mathrm{J\ m^{-3}}]$ | $5.3719\times10^{-16}$ |
| $h[\mathrm{J\ m^{-3}}]$ | $7.2666\times10^{-16}$ |
| $s[\mathrm{J\ m^{-3}\ K^{-1}}]$ | $3.9279\times10^{-17}$ |
| $P[\mathrm{N\ m^{-2}}]$ | $1.8946\times10^{-16}$ |
| $c_V\ [\mathrm{J\ m^{-3}\ K^{-1}}]$ | $9.1501\times10^{-17}$ |
| $n$ | $4.8485\times10^{-28}$ |

**Table 1** Calculated values of the radiative and thermodynamic state functions at redshift $z = 0$ for the average spectrum of the extragalactic far infrared background radiation. $\nu_1 = 0.15\ \mathrm{THz}$, $\nu_2 = 2.4\ \mathrm{THz}$ and $T = 18.5\ \mathrm{K}$.

| **Quantity** | **Extragalactic Average Spectrum** $\nu_1 \le \nu \le \nu_2$ |
| --- | --- |
| $I_0(\nu_1,\nu_2,T)$ $[\mathrm{J\ m^{-3}}]$ | $6.2852\times10^{-15}$ |
| $I_0'(\nu_1,\nu_2,T)$ $[\mathrm{W\ m^{-2}}]$ | $4.7106\times10^{-7}$ |
| $\varepsilon(T)$ | $2.7191\times10^{-6}$ |
| $f\ [\mathrm{J\ m^{-3}}]$ | $-2.0951\times10^{-15}$ |
| $u\ [\mathrm{J\ m^{-3}}]$ | $8.9376\times10^{-15}$ |
| $h[\mathrm{J\ m^{-3}}]$ | $1.1032\times10^{-14}$ |
| $s[\mathrm{J\ m^{-3}\ K^{-1}}]$ | $2.3854\times10^{-16}$ |
| $P[\mathrm{N\ m^{-2}}]$ | $2.0951\times10^{-15}$ |
| $c_V\ [\mathrm{J\ m^{-3}\ K^{-1}}]$ | $9.5092\times10^{-16}$ |
| $n$ | $1.4081\times10^{-27}$ |

**Table 2** Calculated values of the radiative and thermodynamic state functions at redshift $z$ = 1.5 for the average spectrum of the extragalactic far infrared background radiation. $\nu_1 = 2.625\ \mathrm{THz}$, $\nu_2 = 42\ \mathrm{THz}$ and $T = 46.25\ \mathrm{K}$.

| **Quantity** | **Monopole** $0 \le \tilde{\nu} \le \infty$ | **Monopole** $\nu_1 \le \nu \le \nu_2$ | **Dipole** $\nu_1 \le \nu \le \nu_2$ | **Dipole** $0 \le \tilde{\nu} \le \infty$ |
|---|---|---|---|---|
| $a'$, $a''$ $\left[\mathrm{J\ m^{-3}\ K^{-4}}\right]$ | $7.5657\times10^{-16}$ | $7.2170\times10^{-16}$ | $2.9260\times10^{-15}$ | $3.0263\times10^{-15}$ |
| $\sigma'$, $\sigma''$ $\left[\mathrm{W\ m^{-3}\ K^{-4}}\right]$ | $5.6704\times10^{-8}$ | $5.4090\times10^{-8}$ | $2.1930\times10^{-7}$ | $2.2681\times10^{-7}$ |
| $I_0^{\mathrm{M}}(\nu_1,\nu_2,T)$<br>$I_0^{\mathrm{D}}(\nu_1,\nu_2,T)$<br>$\left[\mathrm{J\ m^{-3}}\right]$ | $5.9149\times10^{-2}$<br>- | $5.6422\times10^{-2}$<br>- | -<br>$2.8334\times10^{-4}$ | -<br>$2.9201\times10^{-4}$ |
| $I_0'^{\mathrm{M}}(\nu_1,\nu_2,T)$<br>$I_0'^{\mathrm{D}}(\nu_1,\nu_2,T)$<br>$\left[\mathrm{W\ m^{-2}}\right]$ | $4.4331\times10^{6}$<br>- | $4.2287\times10^{6}$<br>- | -<br>$2.1161\times10^{4}$ | -<br>$2.1885\times10^{4}$ |
| $f\ \left[\mathrm{J\ m^{-3}}\right]$ | $-1.9716\times10^{-2}$ | $-1.8807\times10^{-2}$ | $-9.4112\times10^{-5}$ | $-9.7337\times10^{-5}$ |
| $u\ \left[\mathrm{J\ m^{-3}}\right]$ | $9.8581\times10^{-2}$ | $9.5060\times10^{-2}$ | $5.3819\times10^{-4}$ | $5.8402\times10^{-4}$ |
| $h\left[\mathrm{J\ m^{-3}}\right]$ | $7.8865\times10^{-2}$ | $7.6253\times10^{-2}$ | $2.7351\times10^{-4}$ | $2.9201\times10^{-4}$ |
| $s\left[\mathrm{J\ m^{-3}\ K^{-1}}\right]$ | $2.6522\times10^{-5}$ | $2.5643\times10^{-5}$ | $9.1981\times10^{-8}$ | $9.8203\times10^{-8}$ |
| $P\left[\mathrm{N\ m^{-2}}\right]$ | $1.9716\times10^{-2}$ | $1.8807\times10^{-2}$ | $9.4112\times10^{-5}$ | $9.7337\times10^{-5}$ |
| $c_V\ \left[\mathrm{J\ m^{-3}\ K^{-1}}\right]$ | $7.9567\times10^{-5}$ | $2.4842\times10^{-5}$ | $4.3798\times10^{-4}$ | $5.8402\times10^{-4}$ |
| $n$ | $5.3338\times10^{17}$ | $4.4691\times10^{17}$ | $1.7844\times10^{15}$ | $1.9749\times10^{15}$ |

**Table 3** Calculated values of the radiative and thermodynamic state functions for the monopole and dipole spectra in the 65.4 – 654 THz frequency interval at redshift $z \approx 1089$. $T = 2973.52\ \mathrm{K}$ and $T_{\mathrm{amp}} = 3.67\ \mathrm{K}$.